\documentclass[
prb,
reprint,
superscriptaddress,
showpacs,
amsmath,amssymb,
floatfix,
]{revtex4-1}
\usepackage{dcolumn}% Align table columns on decimal point
\usepackage{bm}% bold math
\usepackage{turnstile}

\usepackage{graphicx}
\usepackage{amsmath,amssymb}
\usepackage{url}
\usepackage{multirow}
\usepackage{float}

\DeclareMathOperator{\sgn}{\mathop{\mathrm{sgn}}}
\DeclareMathOperator{\re}{\mathop{\mathrm{Re}}}

\newcommand{\Eq}[1]{Eq.~(\ref{#1})}
\newcommand{\Eqs}[1]{Eqs.~(\ref{#1})}
\begin{document}

\title{
Competitive 0 and $\pi$ states in  S/F multilayers: multimode approach
}
\author{T. ~Karabassov}
\affiliation{National Research University Higher School of Economics, 101000 Moscow, Russia}
\date{\today}
\author{V.~S.~Stolyarov}
\affiliation{Moscow Institute of Physics and Technology, 141700 Dolgoprudny, Russia}
\affiliation{Dukhov Research Institute of Automatics (VNIIA), 127055 Moscow, Russia}
\author{A.~A.~Golubov}
\affiliation{Faculty of Science and Technology and MESA$^+$ Institute for Nanotechnology,
	University of Twente, 7500 AE Enschede, The Netherlands}
\affiliation{Moscow Institute of Physics and Technology, 141700 Dolgoprudny, Russia}
\author{V.~M.~Silkin}
\affiliation{Donostia International Physics Center (DIPC), Paseo Manuel de Lardizabal 4, San Sebasti\'{a}n/Donostia, 20018 Basque Country, Spain}
\affiliation{Departamento de F\'{\i}sica de Materiales, Facultad de Ciencias Qu\'{\i}micas,
  UPV/EHU, 20080 San Sebasti\'{a}n, Basque Country, Spain}
\affiliation{IKERBASQUE, Basque Foundation for Science, 48011 Bilbao, Spain}
\author{V.~M.~Bayazitov}
\affiliation{N.S. Kurnakov Institute of General and Inorganic Chemistry, Russian Academy of Sciences, 117901 Moscow, Russia}
\author{B.~G.~Lvov}
\affiliation{National Research University Higher School of Economics, 101000 Moscow, Russia}
\author{A.~S.~Vasenko}
\affiliation{National Research University Higher School of Economics, 101000 Moscow, Russia}
\affiliation{I.E. Tamm Department of Theoretical Physics, P.N. Lebedev Physical Institute, Russian Academy of Sciences, 119991 Moscow, Russia}
\begin{abstract}
We have investigated the critical temperature behavior in periodic superconductor/ ferromagnet (S/F) multilayers as a function of the ferromagnetic layer thickness $d_f$ and the interface transparency. The critical temperature $T_c(d_f)$ exhibits a damped oscillatory behavior in these systems due to an exchange field in the ferromagnetic material. In this work we have performed $T_c$ calculations using the self-consistent \textit{multimode approach}, which is considered to be exact solving method. Using this approach we have derived the conditions of 0 or $\pi$ state realization in periodic S/F multilayers. Moreover, we have presented the comparison between the single-mode and multimode approaches and established the limits of applicability of the single-mode approximation, frequently used by experimentalists.
\end{abstract}

\pacs{74.25.F-, 74.45.+c, 74.78.Fk}

\maketitle

%%%%%%%%%%%%%%%%%%%%%%%%%%%%%%%%%%%%%%%%%%%%%%%%%%%%%%%%%%%%%%%%%%%%%%%%%%%%
%%%%%%%%%%%%%%%%%%%%%%%%%%%%%%%%%%%%%%%%%%%%%%%%%%%%%%%%%%%%%%%%%%%%%%%%%%%%
\section{Introduction}

Nowadays, the rates of development of such areas as spintronics and superconducting logic and memory circuits increase significantly. In particular, much attention was attracted to superconductor/ ferromagnet (S/F) structures.\cite{BuzdinRMP, GolubovRMP, BergeretRMP} It is known that S/F structures are important for RSFQ circuits,\cite{Hilgenkamp2008} applications for superconducting spintronics and in particular memory elements, \cite{Tagirov1999, Larkin2012, Linder2015, Golovchanskiy2016, Bakurskiy2016, Soloviev2017, Caruso2018, Golovchanskiy2018_1, Bakurskiy2018}, magnetoelectronics,\cite{Baek2014, Gingrich2016, Golovchanskiy2018_2} qubits,\cite{Feofanov2010} artificial neural networks,\cite{Soloviev2018} micro-refrigerators,\cite{Ozaeta2012, Kawabata2013} etc. It was suggested that arrays of S/F structures may provide new functionality.

Rich physics of S/F systems is based on the proximity effect in S/F bilayers.\cite{BuzdinRMP, GolubovRMP, BergeretRMP, Demler1997, Ozaeta2012R} It turns out that when a superconductor and a ferromagnet form a hybrid structure, superconducting correlations leak into a ferromagnetic metal over the distance $\xi_h=\sqrt{D_f/h}$, where $D_f$ is the diffusion coefficient and $h$ is the exchange field in the ferromagnetic material.\cite{BuzdinRMP} As a consequence it leads to a damped oscillatory behavior of superconducting correlations in the ferromagnetic layer, with characteristic lengths of decay and oscillations given by $\xi_h$. The realized superconducting phase is then similar to the FFLO (Fulde-Ferrel-Larkin-Ovchinnikov) state.\cite{Fulde1964, Larkin1964}

If a ferromagnetic layer serves as a weak link in a Josephson-type S/F/S structure, there is a possibility of the $\pi$ phase state realization. For F layer thickness $d_f \ll \xi_h$ the pair wave function in the F layer changes little, and the sign of the superconducting pair potential in the S layers remains the same. The phase difference in the S layers is then zero (0 phase state). Increasing the F layer thickness up to $d_f \sim \xi_h$, the pair wave function may cross zero in the center of the F layer with the $\pi$ phase shift ($\pi$ phase state) and different signs of the superconducting pair potential in the S layers. Further increasing of $d_f$ may cause subsequent 0-$\pi$ transitions due to damped oscillatory behavior of the pair potential in the F layer. The existence of the $\pi$ states leads to a number of striking phenomena. For example, the critical current in S/F/S Josephson junctions exhibits a damped oscillatory behavior with increasing F layer thickness. \cite{Buzdin1982, Buzdin1991, Ryazanov2001, Blum2002, Sellier2004, Bauer2004, Bell2005, Born2006, Pepe2006, Oboznov2006, Weides2006, Bannykh2009, Vasenko2008, Kemmler2010, Baker2014, Halterman2014, Loria2015, Bakurskiy2017, Yamashita2017} The $\pi$ state is then characterized by the negative sign of the critical current. Similarly, 0 to $\pi$ transitions can also be observed in oscillations of the density of states in the F layer, \cite{Buzdin2000, Kontos2001, Halterman2004, Vasenko2011} and critical temperature $T_c$ oscillations on the ferromagnetic layer thickness in S/F/S structures.\cite{Jiang1995, Jiang1996, Khusainov1997, Tagirov1998, Khusainov2000, Proshin2001, Izyumov2002, Cirillo2009, Kushnir2011_1, Kushnir2011_2, Samokhvalov2015, Kushnir2016, Vdovichev2016, edSidorenko2018, Khaydukov2018} Influence of magnetization misalignment in F layers on critical temperature in F$_1$/S/F$_2$/(S') structures was also investigated.\cite{Proshin2006, Avdeev2010} Josephson $\pi$ junctions have been proposed as elements of novel superconducting nanoelectronics in many different applications.

The critical temperature have nontrivial behavior even in S/F bilayers.\cite{Fominov2002, Cirillo2005, Zdravkov2006, Zdravkov2010, Zdravkov2011, Mancusi2011} In this case the transition to the $\pi$ phase is impossible, but the commensurability effect between the period of the superconducting correlations oscillation ($\sim \xi_h$) and the F layer thickness leads to a nonmonotonic dependence of $T_c$ on the F layer thickness $d_f$. This reentrant behavior is typical for the high-transparent S/F interfaces, while for medium or low-transparent S/F interfaces the $T_c(d_f)$ function is monotonously decaying.\cite{Fominov2002} Critical temperature in F$_1$/F$_2$/S and F$_1$/S/F$_2$ trilayers (spin-valves) with magnetization misalignment was investigated in Refs.~\onlinecite{Zhu2010, Fominov2010, Wu2012}.

\begin{figure}[tb]
\centering
\includegraphics[width=\columnwidth]{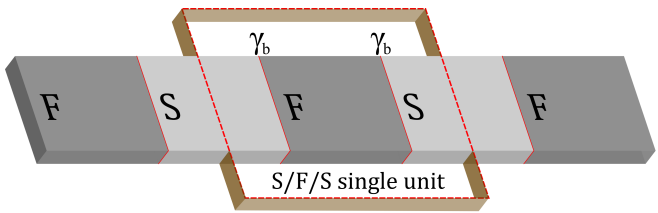}
\caption
{
(Color online) Geometry of the considered S/F multilayer system. Single S/F/S unit is considered in our model. The transparency parameter $\gamma_b$ is proportional to resistance across the S/F interface.
}
\label{SFmult}
\end{figure}

In contrast to bilayers, the S/F periodic multilayered systems may exhibit more complex behavior, with the competition of 0 and $\pi$ phase states.
The purpose of this work is to provide quantitative model of critical temperature $T_c$ calculation in a periodic symmetric S/F multilayer.
Such a system can be divided in S/F/S trilayers as shown in Fig.~\ref{SFmult}.
The total S/F multilayer can be in 0 or in $\pi$ state depending on the state of the single S/F/S unit. It is important to understand the critical temperature behavior of such systems when it comes to practical applications. Therefore it is important to calculate the $T_c(d_f)$ behaviour of the single unit S/F/S trilayer. Previously it was done only in the so called \textit{single mode approximation} (SMA). \cite{Jiang1995, Jiang1996, Khusainov1997, Tagirov1998, Khusainov2000, Proshin2001, Izyumov2002, Cirillo2009, Kushnir2011_1, Kushnir2011_2, Samokhvalov2015, Kushnir2016, Vdovichev2016, edSidorenko2018, Khaydukov2018} In this paper we have calculated the $T_c(d_f)$ dependence, using the \textit{multimode approach} (MMA), considered to be exact method for solving this problem. We also compare the results of the multimode approach with the single mode approximation, setting the limits for the latter approximative method. In our work we do not consider nonequilibrium effects,\cite{Arutyunov2018} and use the Matsubara Green functions technique, which has been developed to describe many-body systems in equilibrium at finite temperature.\cite{Belzig1999}

The paper is organised as follows. In the next section we formulate the theoretical model and basic equations.
In Sections III and IV single mode and multimode approaches are formulated, correspondingly. The results are presented in Section V and summarized and discussed in Section VI.

\section{Model}\label{Sec.model}

%%%%%%%%%%%%%%%%%%%%%%%%%%%%%%%%%%%%%%%%%%%%%%%%%%%%%%%%%%%%%%%%%%%%%%%%%%%%
%%%%%%%%%%%%%%%%%%%%%%%%%%%%%%%%%%%%%%%%%%%%%%%%%%%%%%%%%%%%%%%%%%%%%%%%%%%%
\setcounter{equation}{0}

\begin{figure*}[t]
\centering
\includegraphics[width=2\columnwidth]{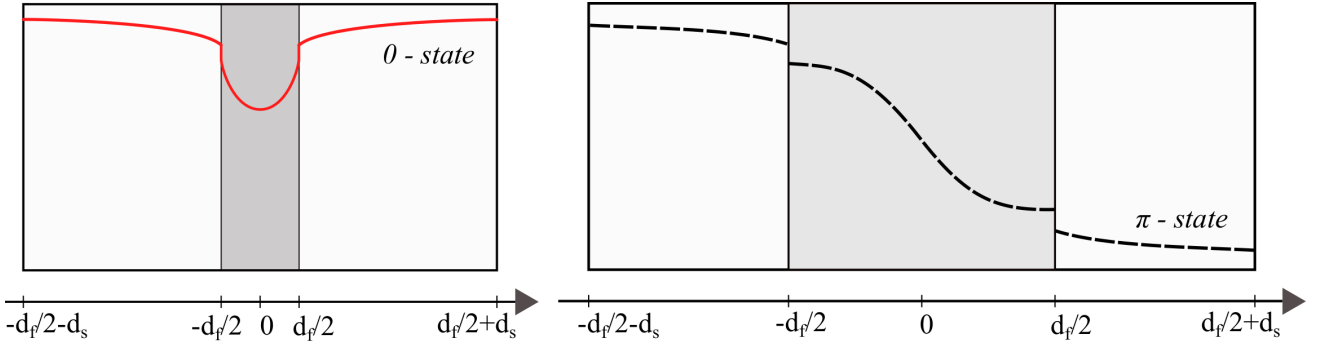}
\caption
{
(Color online) Schematic behavior of the real part of the pair wave function. For thin enough ferromagnetic layer the system is in the 0 phase state (solid red line), while for larger $d_f$ the system can be in the $\pi$ state (dashed black line). Only one of these states is realized depending on the F layer thickness.}
\label{SFS}
\end{figure*}

The model of an S/F/S junction we are going to study is depicted in Fig.~\ref{SFS} and consists of a ferromagnetic layer of
thickness $d_f$ and two superconducting layers of thickness $d_s$ along the $x$ direction. The structure is symmetric and its center is placed at $x=0$.

To calculate the critical temperature $T_c(d_f)$ of this structure we assume the diffusive limit and use the framework of the linearized Usadel equations for the S and F layers in Matsubara representation.\cite{Belzig1999, Usadel1970} Near $T_c$ the normal Green function is $G = \sgn \omega_n$, and the Usadel equation for the anomalous Green function $F$ take the following form. In the S layers ($d_f/2< |x| <d_s + d_f/2$) it reads
\begin{equation}\label{Usadel_S}
\xi_s^2 \pi T_{cs} \frac{d^2 F_s}{d x^2} - |\omega_n| F_s + \Delta = 0.
\end{equation}
In the F layer ($-d_f/2 < x < d_f/2$) the Usadel equation can be written as
\begin{equation}\label{Usadel_F}
\xi_f^2 \pi T_{cs} \frac{d^2 F_f}{d x^2} - \left ( |\omega_n| + i h \sgn \omega_n \right ) F_f = 0.
\end{equation}
Finally, the self-consistency equation reads,\cite{Belzig1999}
\begin{equation}\label{Delta}
\Delta \ln \frac{T_{cs}}{T} = \pi T \sum_{\omega_n} \left ( \frac{\Delta}{|\omega_n|} - F_s \right ).
\end{equation}
In \Eqs{Usadel_S}-\eqref{Delta} $\xi_s = \sqrt{D_s/ 2\pi T_{cs}}$, $\xi_f = \sqrt{D_f/ 2\pi T_{cs}}$, $\omega_n = 2 \pi T (n + \frac{1}{2})$, where $n = 0, \pm 1, \pm 2, \ldots$ are the Matsubara frequencies, $h$ is the exchange field in the ferromagnet, $T_{cs}$ is the critical temperature of the superconductor S, and $F_{s(f)}$ denotes the anomalous Green function in the S(F) region (we assume $\hbar = k_B = 1$). We note that $\xi_h = \xi_f \sqrt{2\pi T_{cs}/h}$.

Equations \eqref{Usadel_S}-\eqref{Delta} should be complemented by the Kupriyanov-Lukichev boundary conditions at the S/F boundaries ($x=\pm d_f/2$),\cite{KL}
\begin{subequations}\label{KL}
\begin{align}
\xi_s \frac{d F_s(\pm d_f/2)}{dx} &= \gamma \xi_f \frac{d F_f(\pm d_f/2)}{dx},
\\
\xi_f \gamma_b \frac{d F_f(\pm d_f/2)}{dx} &= \pm F_s(\pm d_f/2) \mp F_f(\pm d_f/2),
\end{align}
\end{subequations}
where $\gamma = \xi_s \sigma_n / \xi_f \sigma_s$, $\sigma_{s(n)}$ is the normal-state conductivity of the S(F) layer, $\gamma_b = R_B \sigma_n / \xi_f$,\cite{KL, gamma_B} and $R_B$ is the resistance of the S/F boundary (we suppose the symmetric structure with same resistance $R_B$ for $x=\pm  d_f/2$). At the borders of the S layer with a vacuum we naturally have,
\begin{equation}\label{S-vac}
\frac{d F_s( \pm d_s \pm d_f/2)}{dx} = 0.
\end{equation}

The solution of the Usadel equation in the F layer depends on the phase state of the structure. In the 0 phase state the anomalous Green function is symmetric relative to $x=0$ (see Fig.~\ref{SFS}, left panel),\cite{Fominov2002}
\begin{equation}\label{zero}
F_f^0 = C(\omega_n) \cosh \left( k_f x \right),
\end{equation}
while in the $\pi$ phase state the anomalous Green function is antisymmetric (see Fig.~\ref{SFS}, right panel),
\begin{equation}\label{pi}
F_f^\pi = C'(\omega_n) \sinh \left( k_f x \right),
\end{equation}
where
\begin{equation}\label{k_f}
k_f = \frac{1}{\xi_f}\sqrt{\frac{|\omega_n| + i h \sgn \omega_n}{\pi T_{cs}}}.
\end{equation}
In \Eqs{zero},\eqref{pi} the $C(\omega_n)$ and $C' (\omega_n)$ are proportionality coefficients to be found from the boundary conditions.

To solve the boundary value problem \Eqs{Usadel_S}-\eqref{S-vac} we use the method proposed in Ref.~\onlinecite{Fominov2002}.
At the right S/F boundary ($x=d_f/2$) from \Eqs{KL} we obtain,
\begin{align}\label{boundary}
\xi_s \frac{d F_s(d_f/2)}{dx} = \frac{\gamma}{\gamma_b + B_f (\omega_n)} F_s(d_f/2),
\end{align}
where $B_f (\omega_n)$ can acquire one of two different values, depending on phase state. In the 0 phase state,\cite{Fominov2002}
\begin{equation}\label{Bf0}
B_f^0 = \left [ k_f \xi_f \tanh (k_f d_f/2) \right]^{-1},
\end{equation}
while in $\pi$ phase state from \Eq{pi} we obtain,
\begin{equation}\label{Bfpi}
B_f^{\pi} = \left [ k_f \xi_f \coth (k_f d_f/2) \right]^{-1}.
\end{equation}
Similar boundary condition can be written at $x=-d_f/2$.

The boundary condition \eqref{boundary} is complex. In order to rewrite it in a real form, we use the following relation,
\begin{equation}
F^\pm = F(\omega_n) \pm F(-\omega_n).
\end{equation}
According to the Usadel equations \eqref{Usadel_S}-\eqref{Delta}, there is a symmetry relation $F(-\omega_n) = F^*(\omega_n)$ which implies that
$F^+$ is a real while $F^-$ is a purely imaginary function.

Thus we can consider only positive Matsubara frequencies and express the self-consistency equation \eqref{Delta} via the symmetric function $F_s^+$,
\begin{equation}\label{Delta+}
\Delta \ln \frac{T_{cs}}{T} = \pi T \sum_{\omega_n > 0} \left ( \frac{2\Delta}{\omega_n} - F_s^+ \right).
\end{equation}
The problem of determining $T_c$ can be then formulated in a closed form with respect to $F_s^+$. Using the boundary condition \eqref{boundary} we arrive at the effective boundary conditions for $F_s^+$ at the boundaries of the right S layer,
\begin{subequations}\label{B1}
\begin{align}
\xi_s \frac{d F^+_s(d_f/2)}{dx} &= W^{0,\pi}(\omega_n) F^+_s(d_f/2),\label{B1a}
\\
\frac{d F_s^+ (d_s+d_f/2)}{dx} &= 0,\label{B1b}
\end{align}
\end{subequations}
where we used the notations,
\begin{align}
W^{0,\pi}(\omega_n) &= \gamma \frac{A_s \left (\gamma_b + \re B^{0,\pi}_f \right ) + \gamma}{A_s |\gamma_b + B^{0,\pi}_f|^2 + \gamma (\gamma_b + \re B^{0,\pi}_f)},
\\
A_s &= k_s \xi_s \tanh (k_s d_s), \quad k_s = \frac{1}{\xi_s} \sqrt{\frac{\omega_n}{\pi T_{cs}}}.\nonumber
\end{align}
Similar boundary conditions can be written at the boundaries of the left S layer.

The self-consistency equation \eqref{Delta+} and boundary conditions \eqref{B1}, together with the Usadel equation for $F^+_s$,
\begin{equation}\label{Usadel+}
\xi_s^2 \pi T_{cs} \frac{d^2 F_s^+}{d x^2} - \omega_n F_s^+ + 2\Delta = 0
\end{equation}
will be used for finding the critical temperature of the S/F/S structure both in 0 and $\pi$ phase states. In general, this problem should be solved numerically.
%

%%%%%%%%%%%%%%%%%%%%%%%%%%%%%%%%%%%%%%%%%%%%%%%%%%%%%%%%%%%%%%%%%%%%%%%%%%%%%%%%
%%%%%%%%%%%%%%%%%%%%%%%%%%%%%%%%%%%%%%%%%%%%%%%%%%%%%%%%%%%%%%%%%%%%%%%%%%%%%%%%

\section{Single-mode approximation}

In this section we present the single mode approximation (SMA) method.  The solution of the problem \Eqs{B1}-\eqref{Usadel+} can be searched in the form of the following anzatz,
\begin{subequations}\label{Fssingle}
\begin{equation}
F_s^+(x,\omega_n)=f(\omega_n) \cos\left(\Omega\frac{x-d_s-d_f/2}{\xi_s}\right),
\end{equation}
\begin{equation}
\Delta(x)=\delta \cos \left(\Omega\frac{x-d_s-d_f/2}{\xi_s}\right),
\end{equation}
\end{subequations}
where $\delta$ and $\Omega$ do not depend on $\omega_n$. The above solution automatically satisfies boundary condition \eqref{B1b} at $x=d_s+d_f/2$.
Substituting expression \eqref{Fssingle} into the \Eq{Usadel+} we obtain,
\begin{align}\label{f_om}
f(\omega_n)=\frac{2 \delta}{\omega_n + \Omega^2 \pi T_{cs}}.
\end{align}

To determine the critical temperature $T_c$ we have to substitute the \Eqs{Fssingle}-\eqref{f_om} into the self-consistency equation \eqref{Delta+} at $T = T_c$. Then it is possible to rewrite the self-consistency equation in the following form,
\begin{equation}\label{last1}
\ln \frac{T_{cs}}{T_c} = \psi \left ( \frac{1}{2} + \frac{\Omega^2}{2}\frac{T_{cs}}{T_c} \right ) - \psi \left ( \frac{1}{2} \right ),
\end{equation}
where $\psi$ is the digamma function.

Boundary condition \eqref{B1a} at $x=d_f/2$ yields the following equation for $\Omega$,
\begin{equation}\label{last2}
\Omega \tan \left ( \Omega \frac{d_s}{\xi_s} \right ) = W^{0, \pi} (\omega_n).
\end{equation}
The critical temperature $T_c$ is determined by equations \eqref{last1} and \eqref{last2} for both 0 and $\pi$ phase states. These equations extends the model of Ref.~\onlinecite{Fominov2002}, taking into account the possibility of $\pi$ phase state realization in the considered structure.

\section{Multimode approach}

%%%%%%%%%%%%%%%%%%%%%%%%%%%%%%%%%%%%%%%%%%%%%%%%%%%%%%%%%%%%%%%%%%%%%%%%%%%%
%%%%%%%%%%%%%%%%%%%%%%%%%%%%%%%%%%%%%%%%%%%%%%%%%%%%%%%%%%%%%%%%%%%%%%%%%%%%

The multimode approach (MMA) was applied for the first time considering the problem of $T_c$ in an S/N bilayer.\cite{GKLO} Here and below we use
similar notations to that of Ref.~\onlinecite{Fominov2002}.
The solution of the problem \Eqs{B1}-\eqref{Usadel+} within the multimode approach reduces to the equation
\begin{equation}\label{detk}
\det \hat{K}^{0,\pi}=0,
\end{equation}
where the $\hat{K}$ matrix is defined as,
\begin{subequations}
\begin{align}
K_{n0}^{0,\pi} &= \frac{W^{0,\pi}(\omega_n)\cos\left(\Omega_0 d_s/\xi_s\right)- \Omega_0 \sin\left(\Omega_0 d_s/\xi_s\right)}{\omega_n/\pi T_{cs} + \Omega_0^2},
\\
K_{nm}^{0,\pi} &= \frac{W^{0,\pi}(\omega_n)+ \Omega_m \tanh\left(\Omega_m d_s/\xi_s\right)}{\omega_n/\pi T_{cs} - \Omega_m^2},
\end{align}
\end{subequations}
where $n=0,1,...,N$, $m=1,2,...,M$ (we take $M=N$), and $\Omega_n$ are determined by the following equation, obtained from \Eq{Delta+} at $T = T_c$,
\begin{equation}
\ln \frac{T_{cs}}{T_c} = \psi \left ( \frac{1}{2} + \frac{\Omega_n^2}{2}\frac{T_{cs}}{T_c} \right ) - \psi \left ( \frac{1}{2} \right ).\label{dig}
\end{equation}

The multimode approach is considered to be much more accurate comparing to the single-mode approximation, and it was shown in previous studies that in some cases SMA and MMA perform significantly different qualitative behavior for 0 phase state junctions in S/F bilayers.\cite{Fominov2002} In the following, using the multimode approach, we provide calculations of critical temperature for various parameters of the S/F/S structure both in 0 and $\pi$ phase states.
\section{Results}

\begin{figure}[t]
	\centering
	\includegraphics[width=\columnwidth]{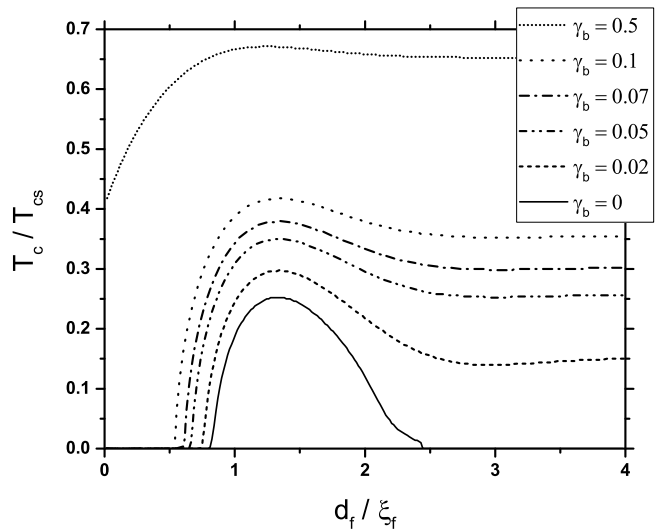}
	\caption
	{$T_c(d_f)$ dependencies for the S/F/S structure in the $\pi$ phase state. $T_c$ is normalized by $T_{cs}$, which is the critical temperature of superconductor in the absence of ferromagnetic layer. We also normalize $d_f$ by oscillation length in the ferromagnetic material $\xi_f$.  Each curve corresponds to particular value of transparency parameter $\gamma_b$. Other parameters are mentioned in the text.}
	\label{MMApial}
\end{figure}

%%%%%%%%%%%%%%%%%%%%%%%%%%%%%%%%%%%%%%%%%%%%%%%%%%%%%%%%%%%%%%%%%%%%%%%%%%%%
%%%%%%%%%%%%%%%%%%%%%%%%%%%%%%%%%%%%%%%%%%%%%%%%%%%%%%%%%%%%%%%%%%%%%%%%%%%%

In this section we present the results obtained by numerical calculations for 0 and $\pi$ phase states using both the single-mode approximation and multimode approach. We provide complete theory for $T_c(d_f)$ behavior description in general case, where system can be in 0 or $\pi$ phase state depending on the F layer thickness $d_f$. Moreover, comparison between the SMA and MMA is also presented. The accuracy of calculations was checked by choosing sufficiently large matrix $\hat{K}$ dimensions in multimode approach. Here and below we have used in our calculations the same parameters as in Ref.~\onlinecite{Fominov2002}, i.e. $\gamma=0.15,  h= 6.8 \pi T_{cs},  d_s= 1.24\xi_s$.

%%%%%%%%%%%%%%%%%%%%%%%%%%%%%%%%%%%%%%%%%%%%%%%%%%%%%%%%%%%%%%%%%%%%%%%%%%%%
\subsection{$T_c$ in S/F/S structures in $\pi$ phase state}

In Fig.~\ref{MMApial} the critical temperature $T_c(d_f)$ dependencies on ferromagnetic layer thickness $d_f$ in the $\pi$ phase state are shown. This situation corresponds to an S/F/S structure enclosed in a ring, where the $\pi$ phase shift can be fixed by applying the magnetic flux quantum for any $d_f$. Different curves correspond to various values of $\gamma_b$ which is proportional to resistance across the S/F interface and can be determined from the experiment. \cite{Mancusi2011} For $\gamma_b=0$ the critical temperature increases to a certain maximum value at a particular $d_f$ and with further rise in thickness $d_f$ the $T_c$ eventually drops to zero. At higher interface resistances $T_c(d_f)$ grows nonmonotonically reaching maximum at finite $d_f$ and then saturating to some value, depending on $\gamma_b$.

\subsection{$T_c$ in S/F/S structures: 0-$\pi$ transitions}

%%%%%%%%%%%%%%%%%%%%%%%%%%%%%%%%%%%%%%%%%%%%%%%%%%%%%%%%%%%%%%%%%%%%%%%%%%%%

\begin{figure}[t]
	\centering
	\includegraphics[width=\columnwidth]{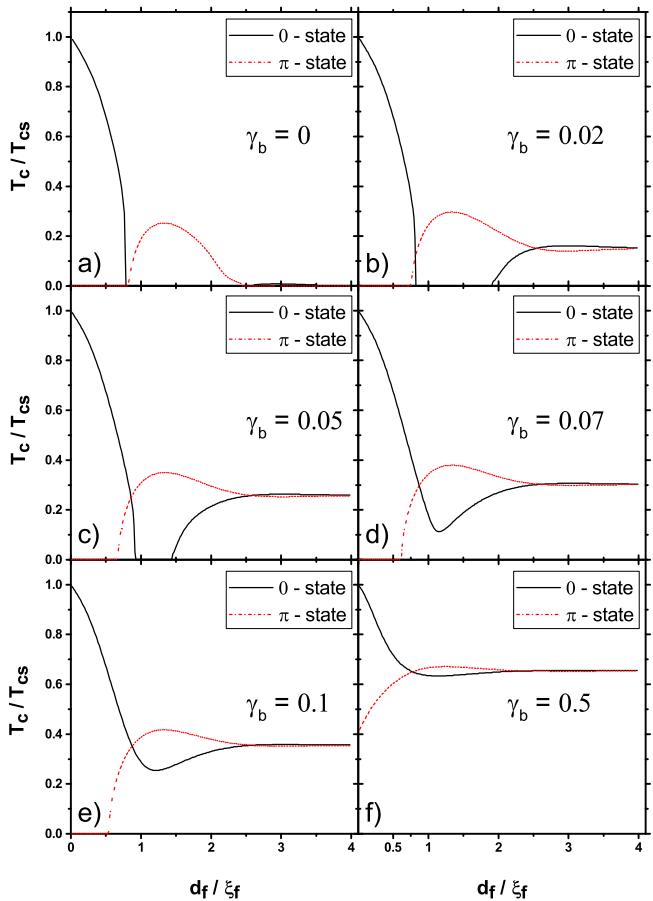}
	\caption
	{(Color online). Plots of $T_c(d_f)$ dependencies in both 0 and $\pi$ phase states. Solid black lines correspond to the 0 phase state, while dashed red lines -  to the $\pi$ phase state. Each plot corresponds to particular value of transparency parameter $\gamma_b$: (a) $\gamma_b=0$, (b) $\gamma_b=0.02$, (c) $\gamma_b = 0.05$, (d) $\gamma_b = 0.07$, (e) $\gamma_b = 0.1$, (f) $\gamma_b = 0.5$.}
	\label{MMApi0}
\end{figure}

\begin{figure}[t]
	\centering
	\includegraphics[width=\columnwidth]{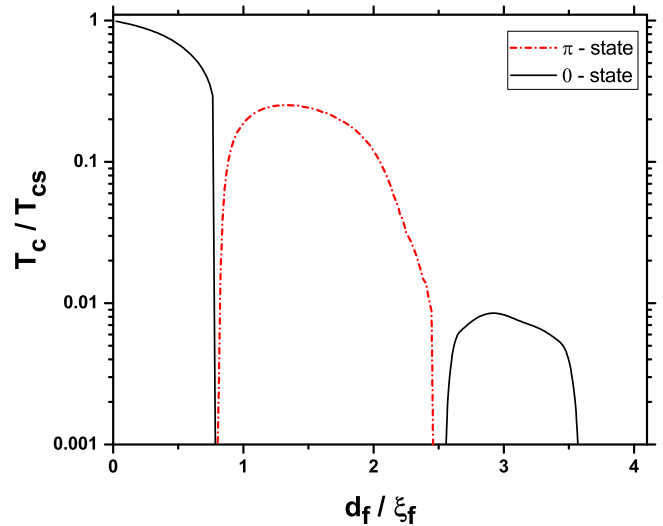}
	\caption
	{(Color online) Illustration of the possibility of multiple 0-$\pi$ transitions in case of $\gamma_b = 0$ (Fig.~\ref{MMApi0}a in logarithmic scale).}
	\label{gamma=0}
\end{figure}

In order to provide complete behavior of the critical temperature in S/F/S systems we calculate $T_c(d_f)$ dependencies in both 0 and $\pi$ phase states and show them on the same plot, see Fig.~\ref{MMApi0}. Both dependencies are calculated for the same set of parameters mentioned above. In S/F/S structures only the state with highest $T_c$ is realized at certain $d_f$, i.e. when increasing $d_f$ the dashed red line lies above the solid black line, the 0-$\pi$ transition occurs and the structure switches to the $\pi$ phase state. Further increasing $d_f$ one can see the $\pi$-0 transition and structure switches back to the 0 phase state. The second transition in case of $\gamma_b = 0$ is shown more clearly in Fig.~\ref{gamma=0} in logarithmic scale. It can be seen that $T_c(d_f)$ has the dumped oscillatory behavior. For all plots we can see that increasing $d_f$ the critical temperature decreases until at some point corresponding to $d_f \sim \xi_h$, $T_c$ starts to grow nonmonotonically and saturate further at higher values of $d_f$ (we note that in our case $\xi_h = 0.54 \xi_f$, since $h = 6.8 \pi T_{cs}$). At the point $d_f \sim \xi_h$ the $\pi$ phase state becomes energetically more favorable. In case of S/F bilayers the $T_c(d_f)$ dependence is described only by 0 state curve and the 0-$\pi$ transitions are impossible. The commensurability effect between $\xi_h$ and $d_f$ nevertheless leads to a nonmonotonic $T_c(d_f)$ dependence in S/F bilayers. \cite{Fominov2002}

Analyzing these results, one can clearly see the difference in $T_c(d_f)$ behavior for different values of $\gamma_b$ parameter. For low values of $\gamma_b$ the $T_c(d_f)$ curve results in highly pronounced dumped oscillations. Increasing $\gamma_b$ the oscillations amplitude decreases.

\subsection{Comparison of SMA and MMA}
%%%%%%%%%%%%%%%%%%%%%%%%%%%%%%%%%%%%%%%%%%%%%%%%%%%%%%%%%%%%%%%%%%%%%%%%%%%%

\begin{figure}[t]
	\centering
	\includegraphics[width=\columnwidth]{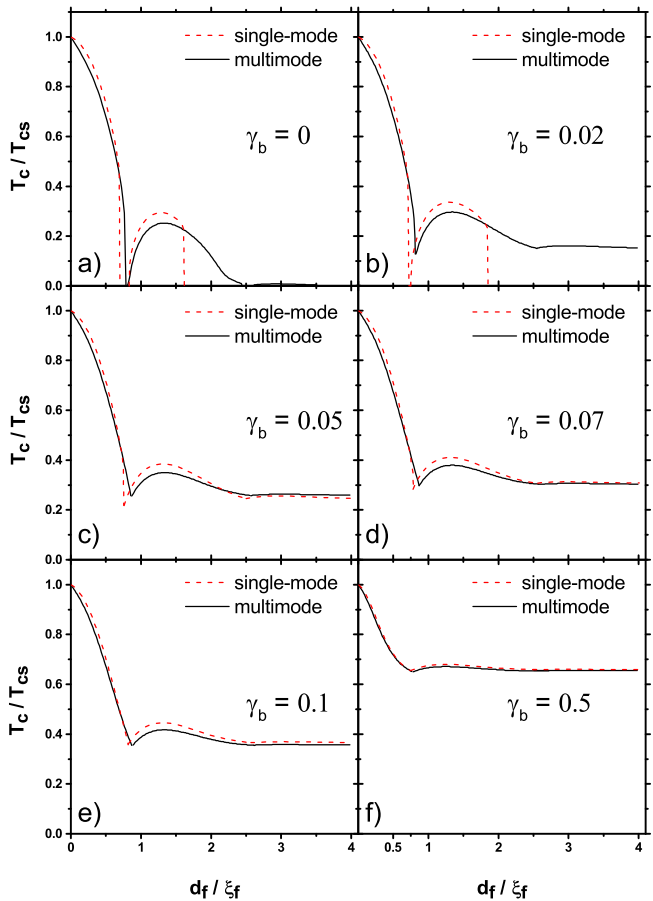}
	\caption
	{
		(Color online) The comparison between single-mode approximation and multimode approach for S/F/S systems. The set of parameters is the same as in Fig.[\ref{MMApial}]. It is clear that for large enough values of $\gamma_b$ both single and multimode approaches perform very close results [(d), (e), and (f)], while there are quantitative and even qualitative differences for small $\gamma_b$ [(a) and (b)].
	}
	\label{SMAallpi}
\end{figure}

The single-mode approximation is frequently used to calculate the critical temperature because of its simplicity. However, that approach has restricted range of applicability which was considered in Ref.~\onlinecite{Fominov2002} for S/F bilayers. Nevertheless the SMA method quite often is used for wide range of parameters, even for values when the approximation can be rude enough, due to its simplicity and speed. The difference between single-mode and multimode methods can be seen in Fig.~\ref{SMAallpi}. For these calculations the same set of parameters as in Fig.~\ref{MMApial} was used. It is important to emphasize, that though for large enough values of $\gamma_b$ both approximations present close and sometimes almost similar results [Fig.~\ref{SMAallpi} (d), (e), (f)], for small enough interface resistance they are quantitatively [Fig.~\ref{SMAallpi} (c)] and even qualitatively different [Fig.~\ref{SMAallpi} (a), (b)]. Using SMA in the latter case the $T_c(d_f)$ jumps to zero abruptly at certain $d_f$, which does not happen in reality. For small enough values of $d_f$ both approximations demonstrate similar results even in this case.

\subsection{Critical thickness of the S layer}

%%%%%%%%%%%%%%%%%%%%%%%%%%%%%%%%%%%%%%%%%%%%%%%%%%%%%%%%%%%%%%%%%%%%%%%%%%%%

\begin{figure}[t]
	\centering
	\includegraphics[width=\columnwidth]{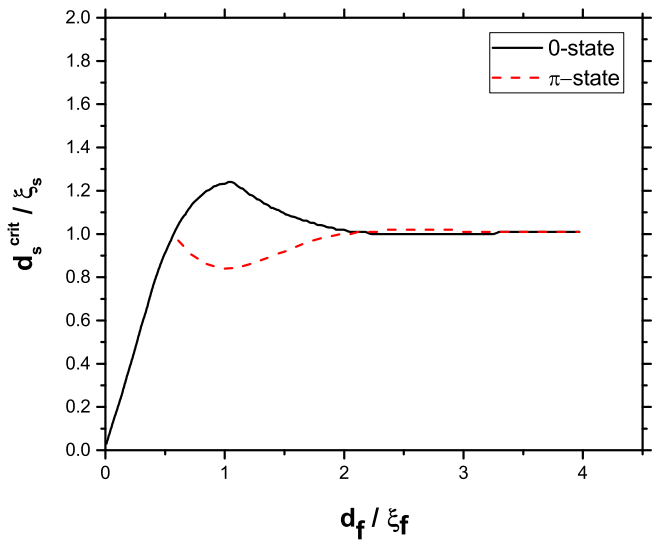}
	\caption
	{
		(Color online) The $d_s^{crit}(d_f)$ dependence calculated by multimode approach. The $\gamma_b=0.1$, other parameters are same as in Fig.~\ref{SMAallpi}.
	}
	\label{dsdf}
\end{figure}

It is well known that decreasing the S layer thickness $d_s$ in S/F multilayer hybrid structures the critical temperature is suppressed since superconductivity is suppressed due to the inverse proximity effect. \cite{BuzdinRMP} Hence at certain value $d_s$ for a given thickness of ferromagnetic layer $d_f$, the critical temperature $T_c$ drops to zero, i.e superconductivity in the structure vanishes. In Fig.~\ref{dsdf} the $d_s^{crit}(d_f)$ dependence is shown both for 0 and $\pi$ phase states of the structure. The S/F/S structure chooses the corresponding phase state (0 or $\pi$) to minimize its energy. Such lowest energy state corresponds to the smallest $d_s^{crit}$. It can be seen from the figure that $d_s^{crit}$ also demonstrates the dumped oscillatory behavior.

%%%%%%%%%%%%%%%%%%%%%%%%%%%%%%%%%%%%%%%%%%%%%%%%%%%%%%%%%%%%%%%%%%%%%%%%%%%%
%%%%%%%%%%%%%%%%%%%%%%%%%%%%%%%%%%%%%%%%%%%%%%%%%%%%%%%%%%%%%%%%%%%%%%%%%%%%

\section{Discussion and conclusion}

%%%%%%%%%%%%%%%%%%%%%%%%%%%%%%%%%%%%%%%%%%%%%%%%%%%%%%%%%%%%%%%%%%%%%%%%%%%%
%%%%%%%%%%%%%%%%%%%%%%%%%%%%%%%%%%%%%%%%%%%%%%%%%%%%%%%%%%%%%%%%%%%%%%%%%%%%

As it was mentioned above the single mode approximation is very popular because of its simplicity and speed. However, this approximation is applicable only in particular range of parameters. The solution of the equations in Sec.~\ref{Sec.model} is accurate when it can be considered as $\omega_n$ independent, which happens in case when $\gamma_b \gg |B_f|$ in \Eq{boundary}. Estimating $|B_f|$ we introduce the real and imaginary parts of $k_f$ in \Eqs{Bf0}-\eqref{Bfpi}, $k_f= k_f' +ik_f''$, and note that $k_f' \gg k_f''$. Then from \Eqs{Bf0}-\eqref{Bfpi} we obtain
\begin{subequations}\label{est}
\begin{align}
|B_f^0| &\sim \left[ k_f' \xi_f \tanh (k_f' d_f) \right]^{-1},\label{est1}
\\
|B_f^\pi| &\sim \left[ k_f' \xi_f \coth (k_f' d_f) \right]^{-1},\label{est1}
\end{align}
\end{subequations}
and finally write the condition $\gamma_b \gg |B_f|$ for 0 phase state,
\begin{equation} \label{condition_0}
\frac 1{\gamma_b} \ll \min\left\{ \sqrt{\max\left( \frac{T_c}{T_{cs}}, \frac{h}{\pi T_{cs}}
	\right)}; \frac{d_f}{\xi_f} \max\left( \frac{T_c}{T_{cs}}, \frac{h}{\pi T_{cs}} \right)
\right\},
\end{equation}
and for $\pi$ phase state,
\begin{equation} \label{condition_pi}
\frac 1{\gamma_b} \ll \min\left\{ \sqrt{\max\left( \frac{T_c}{T_{cs}}, \frac{h}{\pi T_{cs}}
	\right)}; \frac{\xi_f}{d_f} \right\},
\end{equation}
where the ratio $T_c /T_{cs}$ originates from $\omega_n / \pi T_{cs}$ with $\omega_n \sim \pi T_c$ as the characteristic energy scale. The equations \eqref{condition_0}-\eqref{condition_pi} provide the conditions of applicability of single mode approximation both for 0 and $\pi$ phase states.

In this paper we have calculated the $T_c(d_f)$ dependencies for different values of transparency parameter $\gamma_b$. Comparing the results calculated by both SMA and MMA, we note that for high enough resistance across the S/F interfaces, the SMA and MMA show good agreement with each other. However, for highly transparent interfaces, SMA and MMA demonstrate qualitatively and even quantitatively different results. In fact, in this case in the single mode approximation the $T_c$ drops to zero abruptly at particular $d_f$ [see Fig.~\ref{SMAallpi} (a), (b)], while in the multimode approach the superconducting state is still present. Moreover, the $\pi$-0 transition point can be seen using the multimode approach [see Fig.~\ref{gamma=0}]. Thus we confirm the relevance of multimode approach in wide range of parameters in the case of S/F multilayers, where 0-$\pi$ phase transitions are possible. The results of the paper can be extended to the case of S/F$_1$/S/F$_2$ hybrid structures, where F$_1$ and F$_2$ are the ferromagnetic layers with magnetization misalignment.

\begin{acknowledgments}
TK and ASV acknowledge the support of the project T3-93 ``Quantum Solid State Systems'', carried out within the framework of the Basic Research Program at the National Research University Higher School of Economics (HSE) in 2018. VSS acknowledges the support of the Russian Science Foundation (project no 18-72-10118). AAG acknowledges partial support by the EU H2020-WIDESPREAD-05-2017-Twinning project "SPINTECH" under the Grant Agreement No. 810144.
\end{acknowledgments}

%%%%%%%%%%%%%%%%%%%%%%%%%%%%%%%%%%%%%%%%%%%%%%%%%%%%%%%%%%%%%%%%%%%%%%%%%%%%
%%%%%%%%%%%%%%%%%%%%%%%%%%%%%%%%%%%%%%%%%%%%%%%%%%%%%%%%%%%%%%%%%%%%%%%%%%%%

\end{document}